# The magnetic behavior of $Li_2MO_3$ (M=Mn, Ru and Ir) and $Li_2(Mn_{1-x}Ru_x)O_3$

I. Felner[a, *] and I. M. Bradarić[b]

[a]The Racah Institute of Physics, The Hebrew University of Jerusalem, Jerusalem 91904, Israel

[b]The "Vinča" Institute of Nuclear Sciences, Laboratory for Theoretical and Condensed Matter Physics, P.O. Box 522, Belgrade 11001, Yugoslavia.

The present study summerizes magnetic and Mossbauer measurements on ceramic $Li_2MO_3$ M= Mn, Ru and Ir and the mixed $Li_2(Mn_{1-x}Ru_x)O_3$ materials, which show many of the features reflecting to antiferromagnetic ordering or to existence of paramagnetic states. $Li_2IrO_3$ and $Li_2RuO_3$ are paramagnetic down to 5 K. $Li_2(Mn_{1-x}Ru_x)O_3$ compounds are antiferromagnetically ordered at $T_N$ = 48 K for x=0. $T_N$ decreases as the Ru content increases and, for x=0.8, $T_N$ =34 K.

[*]Corresponding author Tel: 972-2-6585752: fax: 972-2-6586347.

E-mail address: israela@vms.huji.ac.il (I.Felner)

**Introduction**

Numerous studies carried out in recent years on ternary ruthenates and ternary manganites have revealed a wide range of electronic and magnetic properties, ranging from superconductivity to ferromagnetism. One class of oxides that has attracted renewed interest, due to their unusual magnetic properties, are the orthorhombic perovskite $CaRuO_3$ compounds [1-2],. The magnetic ground state of $CaRuO_3$ Ru(IV) was established to be *paramagnetic* down to 30 mK[3]. In recent studies we have shown that in ceramic and single-crystalline $Ca(Ru_{1-x}Cu_x)O_3$ samples, irreversibility appears in the zero-field-cooled (ZFC) and field-cooled (FC) curves when measured at low applied magnetic fields [4-5]. A small hysteresis loop opens at low temperatures and Mossbauer studies of 1% $^{57}Fe$ doped in $CaRuO_3$ show a magnetic sextet at 4.1 K which disappears at 90 K. It is proposed that $CaRuO_3$ is not paramagnetic, but rather shows the characteristics of short-range magnetic interactions down to 90 K, possibly as spin-glass like behavior. On the other hand, $CaMnO_3$ is a G-type antiferromagnet (AFM) with $T_N$ around 120 K. The study of the $Ca(Mn_{1-x}Ru_x)O_3$ system shows [6] that ferromagnetism is induced over a large compositional range reaching a maximum value of $T_C$=210 K for x=0.4.

X-ray diffraction studies have shown that metallic $Li_2RuO_3$ Ru (IV) has a monoclinic unit cell [7] (space group C2/c (15) with lattice parameters: a= 5.057 A, b=8.759 A, c=9.854 A and β=100.08 . Magnetic measurements performed down to 80 K, indicate a *paramagnetic* behavior which obeys the Curie-Weiss (CW) law, with an effective moment $P_{eff}$ =1.83 $\mu_B$ and θ =-167 K. On the other hand the isostructural monoclinic compound $Li_2MnO_3$ (a= 4.928 A , b=8.533 A , c=9.604 A and β=99.5 ) is *antiferromagnetically* ordered with $T_N$ around 50 K [8]. The slight decrease in the unit cell volume of $Li_2MnO_3$ is a result of the smaller ionic radius of the Mn (IV) (0.54 A), as compared to 0.63 A for Ru (IV) in $Li_2RuO_3$ [7]. The stark contrast between the two compounds is surprising, because $Li_2MnO_3$ and $Li_2RuO_3$ are closely related, both chemically and structurally. The association of manganese and ruthenium therefore, is of great interest and the investigation of the magnetic behavior of the mixed system $Li_2(Mn_{1-x}Ru_x)O_3$ is the main goal of the present paper.

We present here a comprehensive study of the magnetic properties of ceramic $Li_2(Mn_{1-x}Ru_x)O_3$ measured at various applied fields. Our magnetic studies on $Li_2RuO_3$ measured down to 5 K, as well as Mossbauer studies of 1% $^{57}Fe$ doped in $Li_2RuO_3$ demonstrate the paramagnetic nature of this compound. In addition, we show that $Li_2IrO_3$ is also paramagnetic down to 5 K. On the other hand, $Li_2(Mn_{1-x}Ru_x)O_3$ compounds are AFM ordered and the magnetic ordering temperature decreases continuously (up to x=0.8) with increasing Ru content. This means that the paramagnetic $Li_2RuO_3$ is on the verge of magnetic ordering and readily evolves into a magnetically ordered phase.

**Experimental Details**

Ceramic $Li_2MO_3$ (M= Mn, Ru and Ir) and mixed $Li_2(Mn_{1-x}Ru_x)O_3$ samples were prepared by mixing $Li_2CO_3$ (10% excess) and M metal powders and preheating at 800 C for 24 h. The materials were then pulverized, pelletized and fired again at 800 C for 12 h. The pellets were pulverized and washed in hot distilled water to remove the excess $Li_2CO_3$ and then sintered at 800 C (1000 C for $Li_2RuO_3$) 24h in air. Powder X-ray diffraction (XRD) measurements confirmed the purity of the compounds. Magnetic dc measurements were performed in a Quantum Design superconducting quantum interference device magnetometer (SQUID). A Mossbauer study of $Li_2RuO_3$ sample containing 1% $^{57}Fe$ (doped for Ru) was performed at 4.1 K, using a conventional constant acceleration drive and a 100 mCi $^{57}Co$:Rh source.

**Experimental Results**

XRD studies confirm the monoclinic structure (space-group C2/c), with no secondary phases detected. The XRD patterns and the lattice parameters for all ceramic $Li_2MO_3$ (M= Mn, Ru and Ir) samples are in excellent agreement with Refs. 7,8 and 9. The XRD patterns of the mixed $Li_2(Mn_{1-x}Ru_x)O_3$ samples are similar to their parent compounds and the peaks indexed as 002 reflection (18 < 2θ< 19 ) are shown in Fig. 1. As x increases, the peak position shifts toward smaller angles, consistent with the larger unit cell volume of $Li_2MnO_3$ discussed above. For the x=0.4 and x=0.6 samples, the XRD patterns confirm the monoclinic structure but they do not allow to determine the lattice parameters (the number of peaks is not sufficient).

**$Li_2MO_3$ (M=Ru and Ir)**

Temperature dependent zero-field-cooled (ZFC) and field cooled (FC) magnetization curves M(T) for both $Li_2RuO_3$ and $Li_2IrO_3$ samples have been measured at various applied fields and the data obtained at 10 kOe are shown in Fig. 2. No difference between the two ZFC and FC branches was observed even when measured at low applied fields (10 Oe). $Li_2IrO_3$ is paramagnetic down to 5 K and no anomaly is observed in the M(T) curve. The magnetic properties of $Li_2RuO_3$ are more complex. Fig 2 shows a small cusp around 63 K, which may originate from magnetic ordering.

This cusp appears also in M(T) curves measured at low applied fields. The isothermal magnetization M(H) at 5 K measured up to 50 kOe is linear.

Mossbauer spectroscopy (MS) on Fe-doped materials, has proven to be a powerful tool in the determination of the magnetic nature of the Fe site location. When the ions at this site (Ru) become magnetically ordered, they produce an exchange field at the Fe ions residing in this site. The Fe nuclei experience a magnetic hyperfine field leading to a sextet in the MS spectra. This method has been applied previously to prove the magnetic ordering of $CaRuO_3$ and several other ruthenate-based compounds [4]. The MS spectrum of 1% $^{57}Fe$ diluted in $Li_2RuO_3$ measured at 4.1 K is shown in Fig. 3, and displays a single line with an isomer shift I.S.= 0.45(1) mm/s and a small quadrupole splitting of 0.18(1) mm/s. The presence of a singlet and the absence of a magnetic sextet at 4.1 K are a clear indication of non-magnetically ordered $Fe^{3+}$ ions at the Ru site, and serves as supporting evidence that $Li_2RuO_3$ is paramagnetic at 4.1 K. The origin of the cusp in Fig 2 is now under investigation.

The two curves (even at low fields) have the typical paramagnetic shape and follow closely the Curie-Weiss (CW) law: $\chi = \chi_0 + C/(T-\theta)$, where $\chi_0$ is the temperature independent part of $\chi$, C is the Curie constant, and $\theta$ is the CW temperature. For $Li_2RuO_3$, the fit of the CW law in the range of 75<T<275 K yields: $\chi_0 = -1 \times 10^{-4}$ emu/mol Oe, $\theta = -184(1)$ K, and C= 0.9 emu/mol Oe (Table 1). This corresponds to an effective moment $P_{eff} = 2.68$ $\mu_B$, which is close to the expected 2.83 $\mu_B$ according to Hund`s rule for $Ru^{4+}$ ($4d^4$) in the low spin (S=1) state. Assuming $\chi_0=0$, we obtain $\theta= -171$ K and C=0.83 emu/mol Oe, in excellent agreement with Ref.7. The fit for $Li_2IrO_3$ (see the solid line in Fig. 2) yields: $\chi_0 = -3 \times 10^{-4}$ emu/mol Oe, $\theta = -84(1)$ K, and a Curie constant C= 0.45 emu/mol Oe which corresponds to an effective moment $P_{eff} = 1.9$ $\mu_B$.

**$Li_2(Mn_{1-x}Ru_x)O_3$**

The M(T) curves measured at 10 kOe for all $Li_2(Mn_{1-x}Ru_x)O_3$ compounds are shown in Fig. 4. Here too, the field dependent M(H) plots at 5 K are linear and the ZFC and FC curves (measured at 20 Oe) are identical and no irreversibility is observed. The peaks in Fig. 4 are consistent with an AFM ordering. $T_N$ (determined as the peak position) decreases with increasing the Ru content (Table 1). $T_N$ =48 K obtained for $Li_2MnO_3$, agrees well with Ref. 8, and the peak in $Li_2RuO_3$ (Fig. 2) is much higher than $T_N$ = 34 K for x=0.8. Fitting the M/H(T) data in the paramagnetic range to the modified CW law yields the paramagnetic values given in Table1. The Curie constant obtained for $Li_2MnO_3$ corresponds corresponds to an effective moment $P_{eff} = 3.79$ $\mu_B$. which is close to the expected 3.87 $\mu_B$ according to Hund`s rule for $Mn^{4+}$ ($3d^3$, S=3/2). The C values decrease with increasing Ru content up to x=0.6, and in the limit of uncertainty the x=0.8 and x=1 samples have the same C value (Table 1). Note that all θ values obtained are negative, consistent with the AFM nature of the system.

The nature of the magnetic properties of oxide-ruthenates with narrow 4 d-bands strongly depends on the degree of band filling and bandwidth. $Li_2RuO_3$ is believed to have a narrow itinerant 4 d-band width, composed of Ru $t_{2g}$ and oxygen 2p orbitals, which is too narrow for magnetic ordering. It means that $Li_2RuO_3$ is on the verge of magnetic ordering and readily evolves into a magnetically ordered phase. Indeed, 20% of Mn substitution for Ru induces AFM ordering at T= 34 K.

**Acknowledgment**: We are grateful to Dr. U. Asaf for assistance in the XRD experiments. I. Felner gratefully acknowledges support from the BSF (1999). I. M. Bradarić gratefully

**Figure Captions**

Fig. 1 The (002) XRD peak position for $Li_2(Mn_{1-x}Ru_x)O_3$.

Fig. 2 The magnetization curves of the paramagnetic $Li_2MO_3$ (M=Ru and Ir). Fig. 3 Mossbauer spectrum at 4.1 K for 1% Fe doped in $Li_2RuO_3$.

Fig. 4 The temperature dependence of the magnetization of $Li_2(Mn_{1-x}Ru_x)O_3$ indicating the AFM nature of the system. The solid line is the fit to Curie-Weiss law.

**Table 1**

| Compound | $T_N$ (K) | C(emu/mol Oe) | $\theta$ (K) | $\chi_0$ ($10^{-4}$ emu/mol Oe) |
|---|---|---|---|---|
| $Li_2MnO_3$ | 48 (1) | 1.79 | -44(1) | -4(0.4) |
| $Li_2Mn_{0.8}Ru_{0.2}O_3$ | 43 (1) | 1.66 | -54(1) | -2.7(0.3) |
| $Li_2Mn_{0.6}Ru_{0.4}O_3$ | 41 (1) | 1.46 | -64(1) | -3.2(0.4) |
| $Li_2Mn_{0.4}Ru_{0.6}O_3$ | 37 (1) | 0.87 | -83(1) | -2(0.5) |
| $Li_2Mn_{0.2}Ru_{0.8}O_3$ | 34 (1) | 0.92 | -84(1) | -2(0.5) |
| $Li_2RuO_3$ | | 0.90 | -183(1) | -1 (0.1) |
| | | 0.83 | -171 | - |

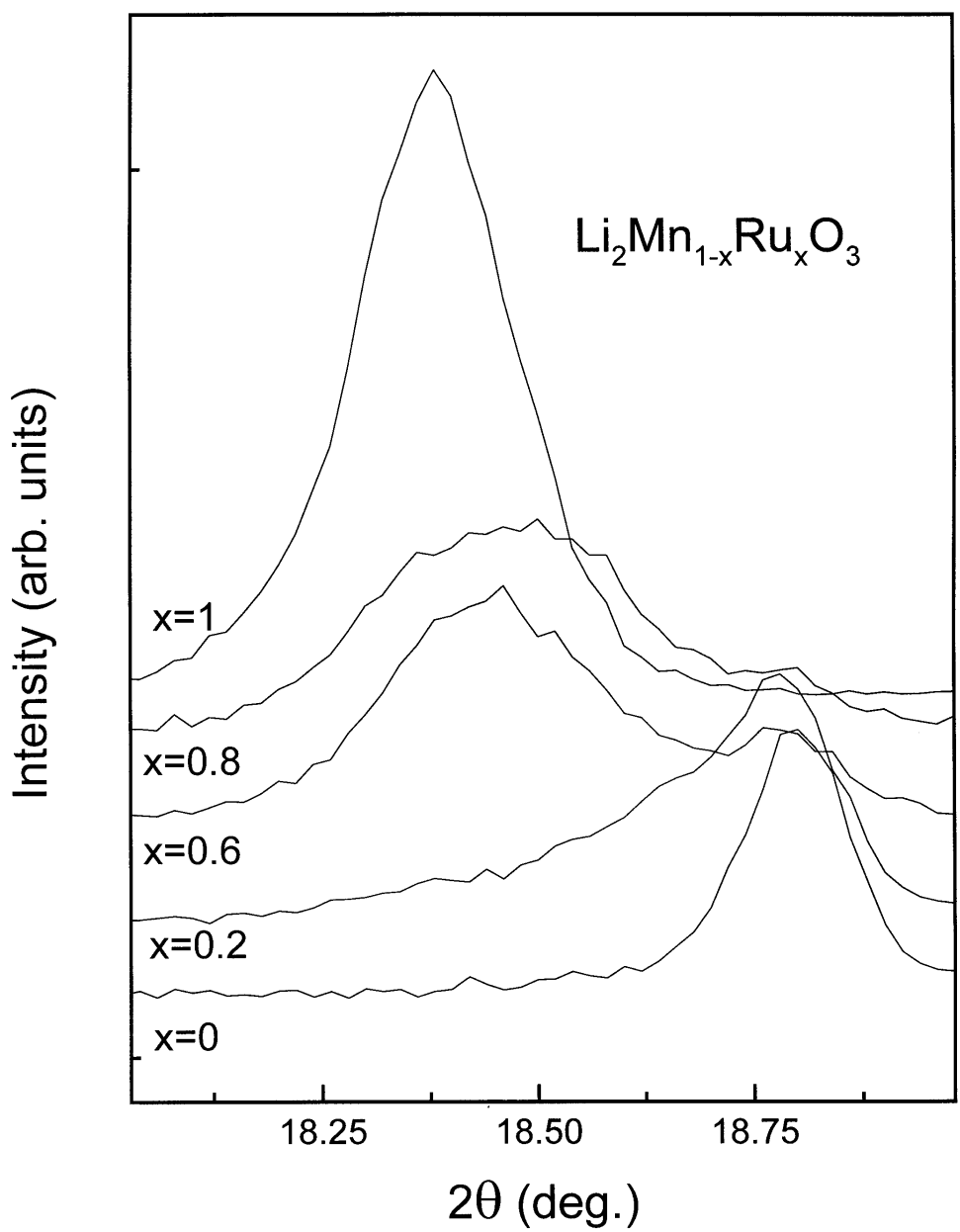

Fig. 1 The (002) XRD peak position for $Li_2(Mn_{1-x}Ru_x)O_3$.

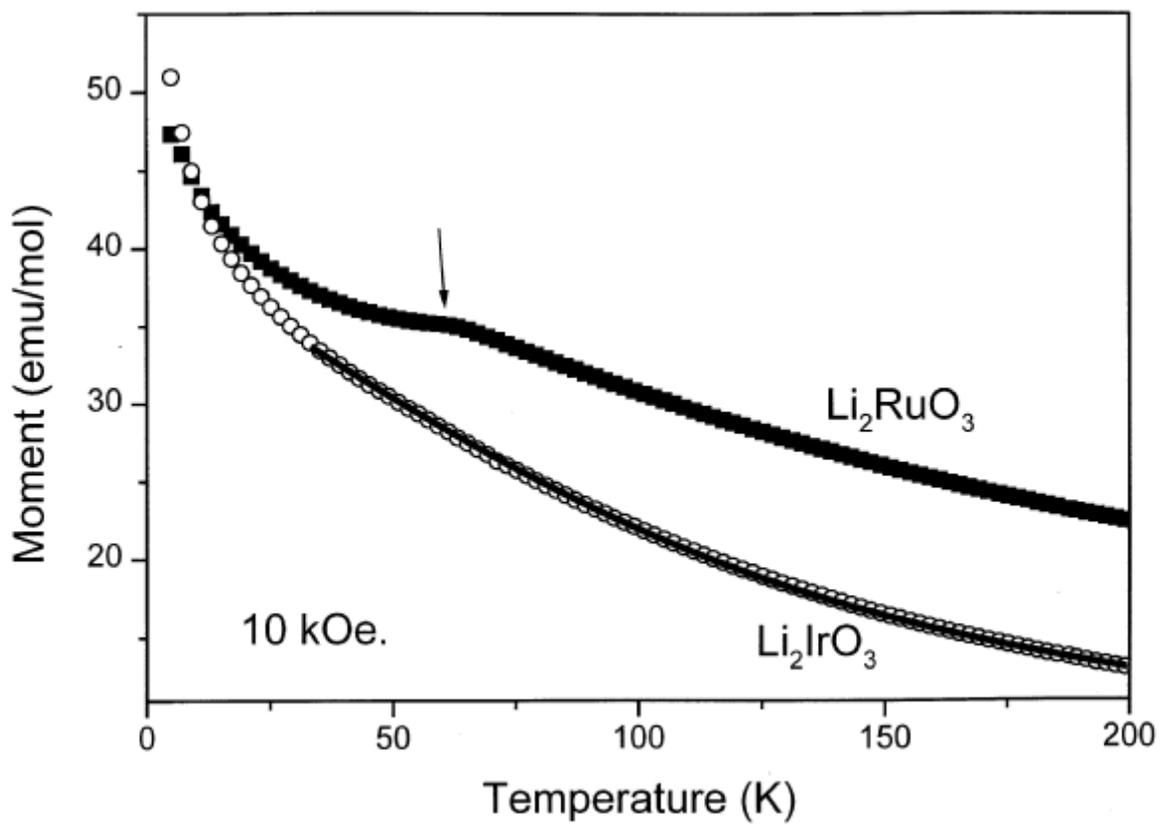

Fig. 2 The magnetization curves of the paramagnetic $Li_2MO_3$ (M=Ru and Ir).

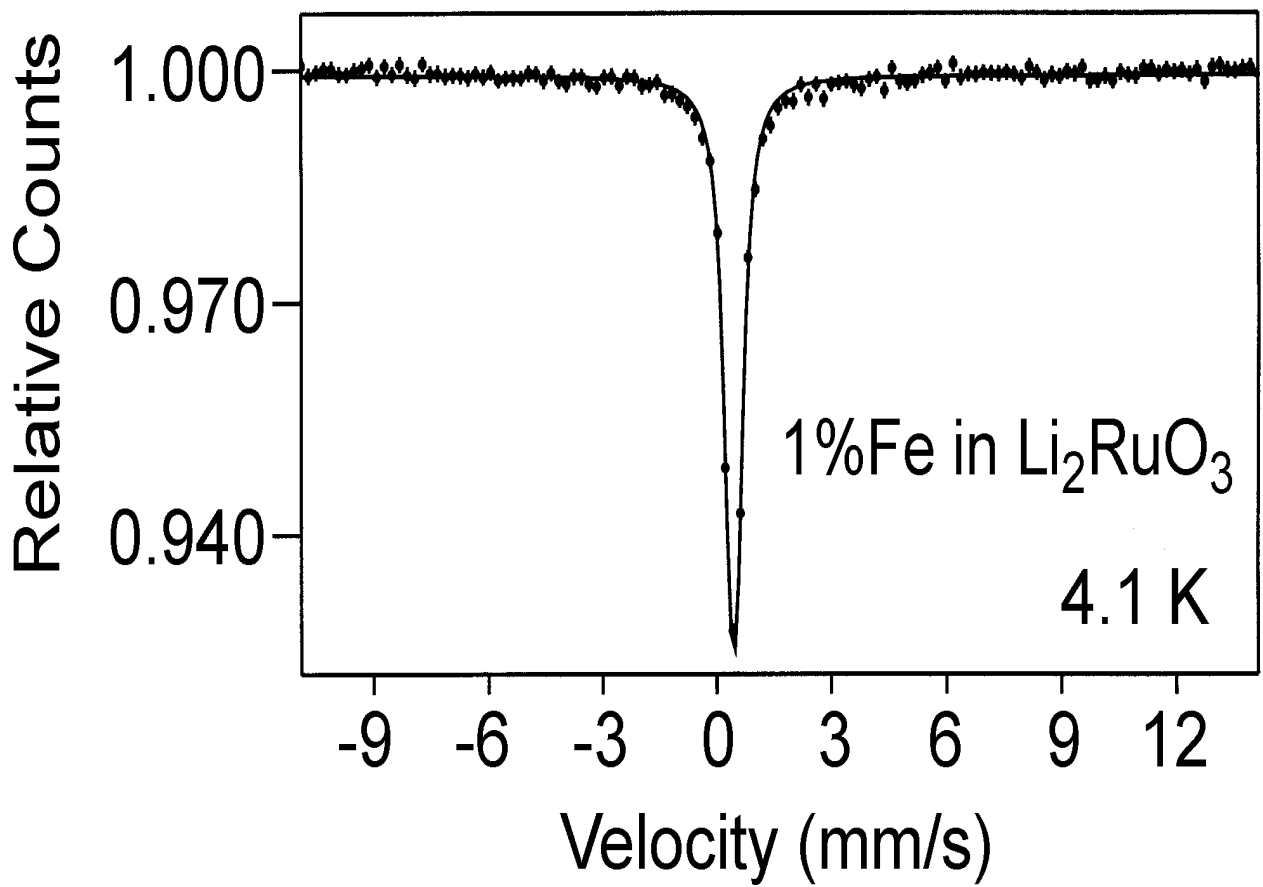

Fig. 3 Mossbauer spectrum at 4.1 K for 1% Fe doped in $Li_2RuO_3$.

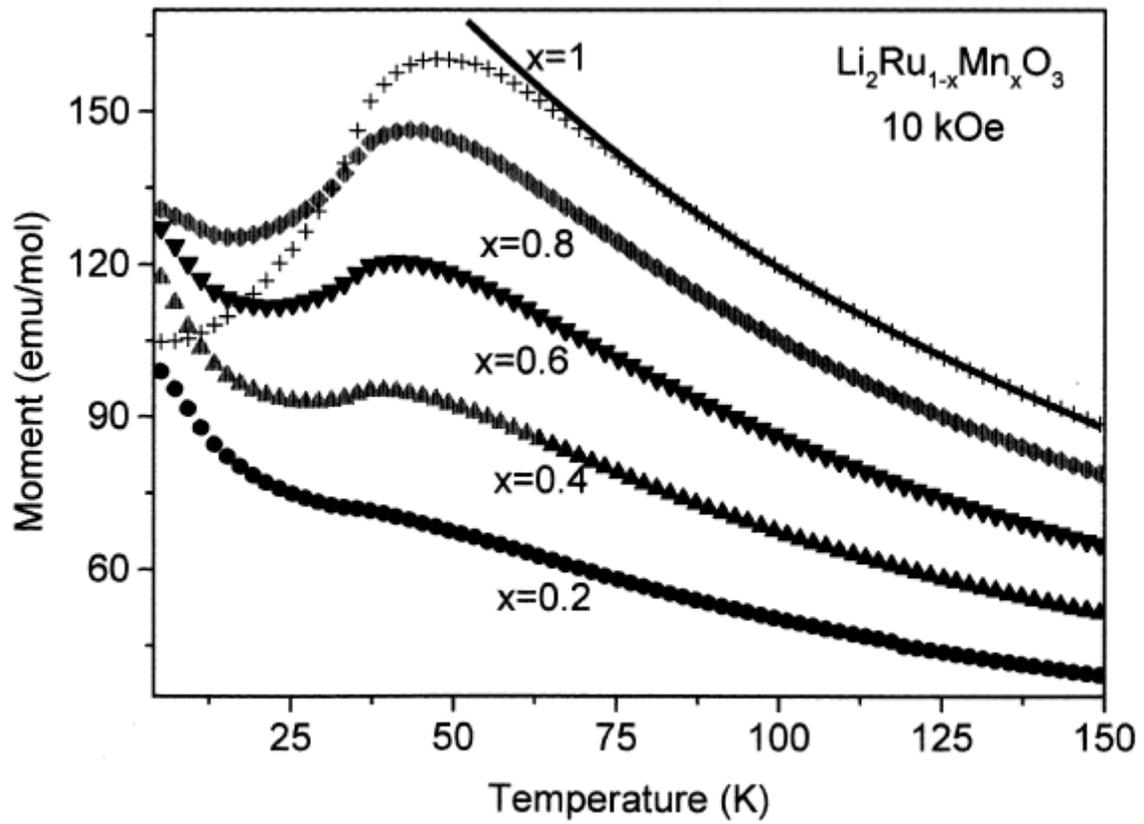

Fig. 4. The temperature dependence of the magnetization of $Li_2(Mn_{1-x}Ru_x)O_3$ indicating the AFM nature of the system.